\documentclass[a4paper,11pt]{article}
\pdfoutput=1 

\usepackage{jinstpub} 
\usepackage{graphicx}
\usepackage{subfig}

\title{Testbeam performance of a shashlik calorimeter with fine-grained longitudinal segmentation}

\newcommand{\nima}        [3]  {{\emph{Nucl.\ Instrum.\ Meth.\ A }}~{\bf #1} (#2) #3}
\newcommand{\itns}        [3]  {{\emph{IEEE Trans. Nucl. Sci.}}~{\bf #1} (#2) #3}
\newcommand{\jinst}       [3]  {{\emph{JINST}}~{\bf #1} (#2) #3}

\author[a,b]{G.~Ballerini,}
\author[a,b]{A.~Berra,}
\author[b,c]{R.~Boanta,}
\author[a,b]{C.~Brizzolari,} 
\author[g]{G.~Brunetti,} 
\author[d]{M.G.~Catanesi,} 
\author[e]{S.~Cecchini,} 
\author[e]{F.~Cindolo,}
\author[b,c]{A.~Coffani,}
\author[f,g]{G.~Collazuol,}
\author[g]{E.~Conti,}
\author[g]{F.~Dal Corso,}
\author[h]{G.~De Rosa,}
\author[i]{A.~Gola,}
\author[l,m]{C.~Jollet,}
\author[g]{A.~Longhin,}
\author[n]{L.~Ludovici,} 
\author[d]{L.~Magaletti,} 
\author[e]{G.~Mandrioli,}
\author[e]{A.~Margotti,}  
\author[a,b]{V.~Mascagna,}
\author[m]{A.~Meregaglia,}
\author[f,g]{M.~Pari,}
\author[e,o]{L.~Pasqualini,}
\author[i]{G.~Paternoster,}
\author[e]{L.~Patrizii,}
\author[i]{C.~Piemonte,} 
\author[e]{M.~Pozzato,} 
\author[g]{F.~Pupilli,}
\author[a,b]{M.~Prest,}
\author[d]{E.~Radicioni,}
\author[h]{A.C.~Ruggeri,}
\author[e]{G.~Sirri,} 
\author[a,b]{M.~Soldani,} 
\author[e]{M.~Tenti,} 
\author[b,c,1]{F.~Terranova, \note{Corresponding author.}} 
\author[p]{E.~Vallazza}

\affiliation[a]{Universit\`a degli Studi dell'Insubria, Via Valleggio 11, Como, Italy}
\affiliation[b]{INFN Milano Bicocca, Piazza della Scienza 3, Milano, Italy}
\affiliation[c]{Universit\`a degli Studi di Milano Bicocca, Piazza della Scienza 3, Milano, Italy}
\affiliation[d]{INFN Sezione di Bari, Via E. Orabona 4, Bari, Italy}
\affiliation[e]{INFN Sezione di Bologna, Via Berti Pichat 6, Bologna, Italy}
\affiliation[f]{Universit\`a degli Studi di Padova, Via Marzolo 8, Padova, Italy}
\affiliation[g]{INFN Padova, Via Marzolo 8, Padova, Italy}
\affiliation[h]{INFN Napoli, Via Cintia, Napoli, Italy}
\affiliation[i]{Fondazione Bruno Kessler, Via Sommarive 18, Povo (TN), Italy}
\affiliation[l]{Institute Pluridisciplinaire Hubert Curien, 23 rue du Loess, Strasbourg, France}
\affiliation[m]{Centre de Etudes Nucleaires de Bordeaux Gradignan, 19 Chemin du Solarium, Bordeaux, France}
\affiliation[n]{INFN Roma, Piazzale Aldo Moro 2, Roma, Italy}
\affiliation[o]{Universit\`a degli Studi di Bologna, Via Irnerio 46, Bologna, Italy}
\affiliation[p]{INFN Trieste, Padriciano 99, 34012 Trieste, Italy}

\emailAdd{francesco.terranova@cern.ch}

\abstract{An iron- plastic-scintillator shashlik calorimeter with a
  4.3~$X_0$ longitudinal segmentation was tested in November 2016 at
  the CERN East Area facility with charged particles up to 5~GeV. The
  performance of this detector in terms of electron energy resolution,
  linearity, response to muons and hadron showers are presented in
  this paper and compared with simulation. Such a
  fine-grained longitudinal segmentation is achieved using a very
  compact light readout system developed by the SCENTT and ENUBET
  Collaborations, which is based on fiber-SiPM coupling boards
  embedded in the bulk of the detector. We demonstrate that this
  system fulfills the requirements for neutrino physics applications
  and discuss performance and additional improvements.}


\begin{document}
\maketitle
\flushbottom

\section{Introduction}
\label{sec:intro}

Shashlik calorimeters~\cite{Fessler:1984wa,atoyan1992} are sampling
calorimeters in which the scintillation light is read-out by
wavelength shifting (WLS) fibers running perpendicularly to the
absorber and converter plates.  Shashlik devices are used since more
than 20 years in particle
physics~\cite{Alvsvaag:1998bd,Aphecetche:2003zr,Zoccoli:2000dn,Goebel:2000uf,Dzhelyadin:2007zz}
but the recent developments in the technology of silicon-based
photosensors provide new solutions for light collection and
readout~\cite{Atoian:2007up,Fantoni:2011zza,Anfimov:2013kka,Berra:2011zz,Berra_shashlik_jinst,jack_maroc}
and open a broader range of applications for shashlik
detectors~\cite{Anelli:2015pba,enubet}. The most critical limitation of these
calorimeters is due to longitudinal segmentation.  Since the light is
transmitted to the photosensors in a matrix of parallel WLS fibers,
bundling and routing of the fibers must be performed at the back of
the calorimeter, preventing longitudinal segmentation. Large area
photosensors directly coupled to the scintillator provide longitudinal
segmentation at the price of introducing dead areas~\cite{Benvenuti:1999qy}. In
calorimetric applications~\cite{Berra:2011zz,Berra_shashlik_jinst,jack_maroc} Silicon PhotoMultipliers
(SiPM) offer performance comparable with conventional PMT but can be
miniaturized down to the scale of a SMD component, whose size can be
matched to the diameter of the WLS fibers~\cite{Berra:2016thx}.

\begin{figure}[htp]
\begin{center}
\subfloat[]{%
  \includegraphics[clip,width=0.8\columnwidth]{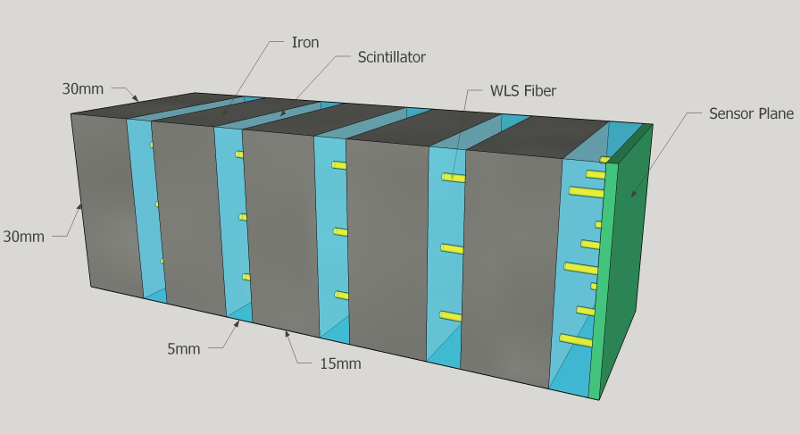}%
} \\
\subfloat[]{%
  \includegraphics[clip,width=0.8\columnwidth]{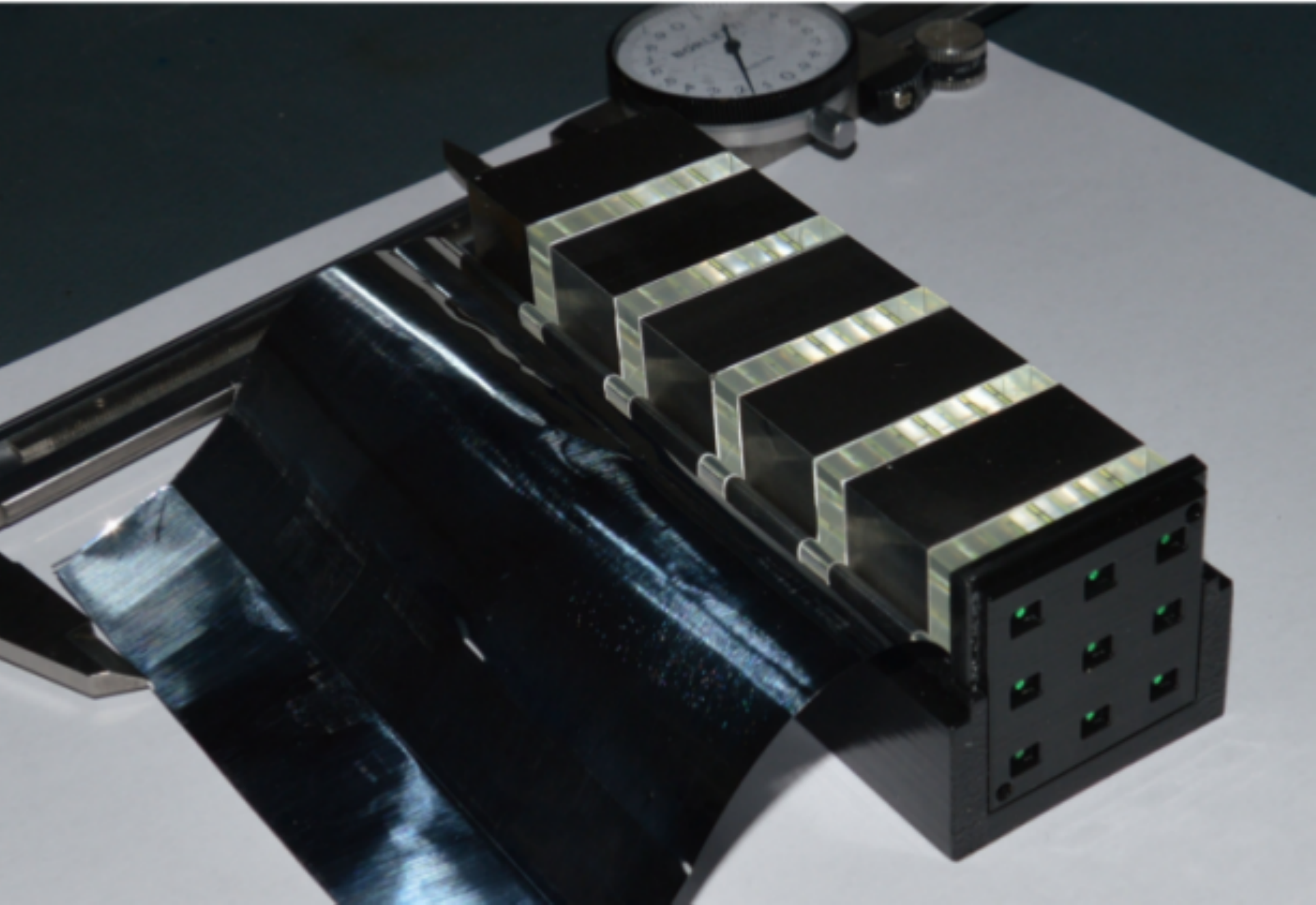}%
}
\end{center}
\caption{ \label{fig:UCM} (a) Layout of the SCENTT ultra-compact module
  (UCM).  The UCM samples 4.3$X_0$ and has a transverse size of
  3$\times$3~cm$^2$. The light is collected by 9 WLS fibers coupled to
  9 SiPMs on the PCB connected to the plastic mask (``sensor
  plane'').  (b) Photo of a UCM used for cosmic ray tests: the PCB is
  not mounted and the plastic mask holding the fibers is visible in
  the back of the detector.  }
\end{figure}

The INFN SCENTT Collaboration has developed an ultra-compact module
(UCM - Fig.~\ref{fig:UCM}) where every single fiber segment is
directly connected to a SiPM. The array of SiPMs reading the UCM is
hosted on a PCB (Printed Circuit Board) holder that integrates both
the passive components and the signal routing toward the front-end
electronics. The calorimeters are assembled grouping arrays of UCMs,
whose size and thickness (in radiation lengths, $X_0$) are optimized
for specific physics applications.

This scheme combines the compactness of SiPM-based
calorimeters~\cite{calice} with the flexibility offered by the
shashlik technology in choosing the longitudinal sampling (length of
the fiber crossing the scintillator/absorber tiles) and transverse
granularity (tile size, number of fibers per unit surface and number
of summed SiPM channels).

The first UCM based calorimeter (``12-module prototype'') was
tested in summer 2016 and results are reported in
Ref.~\cite{test_12module}. The calorimeter tested in November 2016
allows for complete (i.e. transverse and longitudinal) containment of
electromagnetic showers and longitudinal containment of hadronic
showers in the energy range of interest for ENUBET~\cite{enubet}
(1-5~GeV). In ENUBET, UCM-based calorimeters are employed to monitor
positron production in the decay tunnel of conventional neutrino beams
to perform a precise measurement of the $\nu_e$ flux originating from
kaon decays ($K^+ \rightarrow e^+ \pi^0 \nu_e$). For this
application, a longitudinal segmentation of about 4~$X_0$ is needed to
separate positrons from charged pions with a misidentification
probability $<$3\%~\cite{Meregaglia:2016vxf}. The most cost-effective
solution is based on an iron-plastic scintillator UCM with $3\times
3$~cm$^2$ tiles (see Sec.~\ref{sec:detector}).

The calorimeter tested in November 2016 is made of 56 UCMs and a
coarse grained energy catcher located on top of the UCM. The layout
and construction technique employed for this detector are described in
Sec.~\ref{sec:detector}. The calorimeter was characterized at the CERN
East Area (T9 beamline): the experimental area and the ancillary
detectors (silicon trackers, muon catcher, Cerenkov counters) are
described in Sec.~\ref{sec:experiment}. Simulation and equalization of
the response are detailed in Sec.~\ref{sec:simulation} and the detector
response to electromagnetic showers is discussed in
Sec.~\ref{sec:electrons}.  The response to charged pions and
comparisons with simulation are presented in Sec.~\ref{sec:hadrons}.

\section{Layout and construction of the calorimeter}
\label{sec:detector}
 
The calorimeter is made up of 56 UCM modules:
4$\times$2 in the plane perpendicular to the beam and 7 in the
longitudinal direction. On top of this system, an energy tail catcher
(``hadronic module'') is positioned.  The thickness of the iron
absorber is 1.5~cm and the thickness of the scintillator tile is
0.5~cm. The light produced in the scintillator is readout by 9 WLS
fibers (diameter: 1~mm, length:10~cm) and each UCM is composed by five
tiles: it thus samples 4.3~$X_0$ along the development of the shower and 1.7 Moliere radius in the transverse plane.

\begin{figure}[!htb]
\centering
\includegraphics[width=0.9\columnwidth]{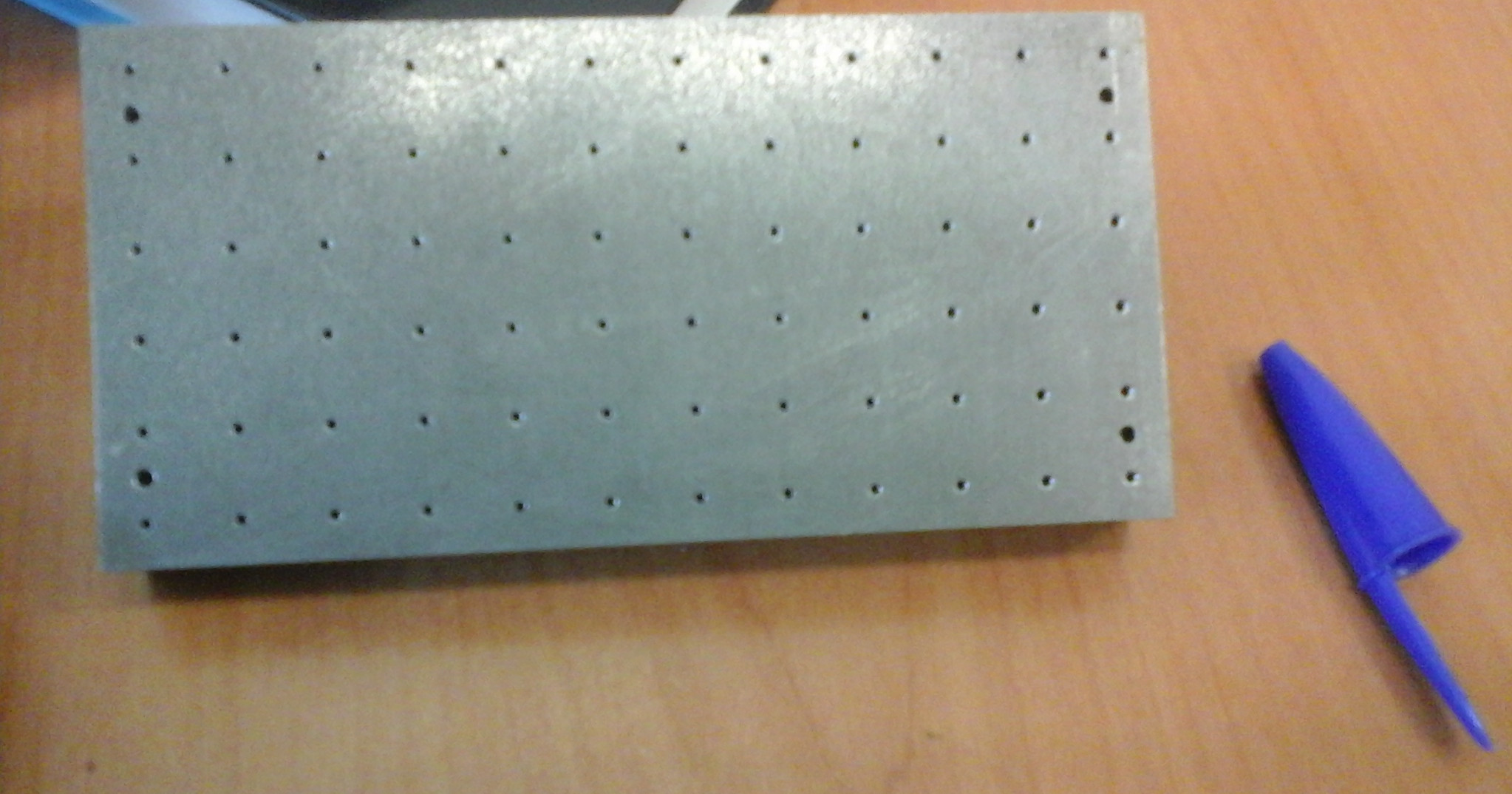}
\caption[]{\label{fig:iron_slab} An iron slab after drilling and
  before zinc coating. The four 2~mm holes for the threaded bolt are
  located in the proximity of the corners.  }
\end{figure}

The UCMs are assembled from 12$\times$6~cm$^2$ iron slabs (standard EN-10025:
S235 -- Fe 360) with 1.5~cm thickness
(Fig.~\ref{fig:iron_slab}). The slabs are drilled with a CNC machine:
the distance between holes is 1~cm and the diameter of the holes is
1.2$\pm$0.2~mm. After drilling, electrolitic zinc-plating is employed
to prevent oxidation.

\begin{figure}[!htb]
\centering
\includegraphics[width=0.9\columnwidth]{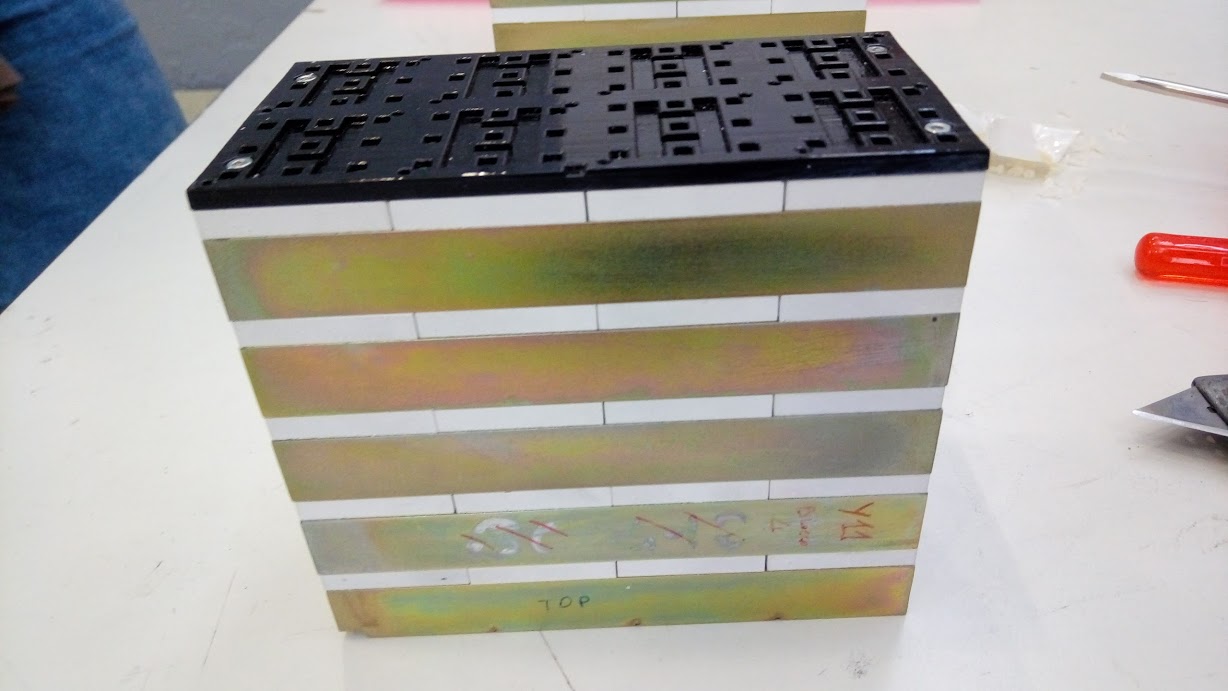}
\caption[]{\label{fig:module} A module of 8 UCMs before the installation
of the PCB. The WLS fibers are hold by the black plastic mask on top. }
\end{figure}

The scintillator tiles (3$\times$3~cm$^2$, thickness 0.5~cm) are
machined using the same procedure employed for the 12-module
prototype~\cite{test_12module}: the tiles were machined and polished
from EJ-200~\cite{eljen} plastic scintillator sheets and painted with
a diffusive TiO$_2$-based coating (EJ-510). Diffusive coating is used
to increase the light collection efficiency and eases the assembly of
the modules compared with more conventional techniques employed in
shashlik devices (e.g. insertion of
Tyvek\textsuperscript{\textregistered} foils between the scintillator
and absorber tiles). The painted scintillators are positioned into the
CNC machine in stacks (up to 4 tiles per stack) and drilled. Each tile
has 9 holes with a diameter of 1.2$\pm 0.1$~mm. A module of 8~UCMs is
assembled from five iron slabs interleaved with planes of scintillator
tiles (8 tiles per slab). The 1~mm diameter WLS fibers cross the
modules through the holes up to the last scintillator plane and are
connected to a plastic mask located downstream of the module, as shown
in Fig.~\ref{fig:module}.

\begin{figure}[!htb]
\centering
\includegraphics[width=0.9\columnwidth]{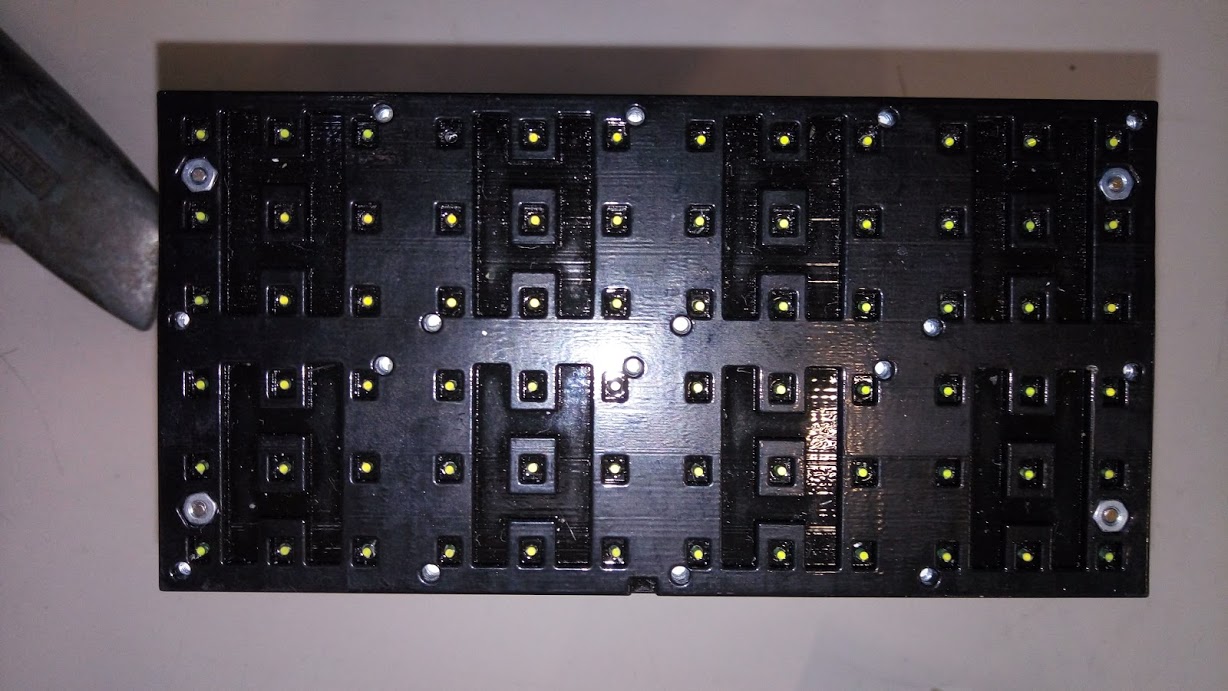}
\caption[]{\label{fig:mask} The plastic mask holding the fibers and the PCB (not shown).
The grooves in the proximity of the fibers host the SiPMs.}
\end{figure}

The plastic mask (see Fig.~\ref{fig:mask}) is built using a 3D printer:
it is grooved in order to fix the fibers in the back of the module and
couple them with the PCB hosting the SiPMs. Four threaded bolts (2~mm
diameter) cross the module and are fixed to the plastic mask by
nuts positioned into the mask itself. The fibers used
for the first six modules are Kuraray~\cite{kuraray} Y11 multi-clad (1~mm
diameter).  The last module is equipped with Saint
Gobain~\cite{saintgobain} BCF92 multi-clad fibers. These fibers offer a fast
response (2.7~ns) compared with Y11 (10~ns) but exhibit a poorer
optical matching to EJ-200, resulting in a 30\% reduction of the light
yield.

The energy tail catcher (hadronic module) is built using the same
technology as for the UCM modules with a coarser granularity. The size of the iron
slabs is 18$\times$9~cm$^2$ (1.5~cm thickness). The hole density
(1~hole/cm$^2$, 1.2~mm diameter) is the same as for the fine-grained
modules. The scintillator tiles of the hadronic module are, however,
much larger: each tile is 18$\times$9~cm$^2$ (thickness 0.5~cm) and
the hadronic module is made of 31 iron slabs interleaved with
scintillator tiles. The tiles are painted and drilled using the same
procedure employed for the fine-grained modules and the light is read
out by BCF92 fibers (1~mm diameter).

\begin{figure}[htp]
\begin{center}
\subfloat[]{%
  \includegraphics[clip,width=0.9\columnwidth]{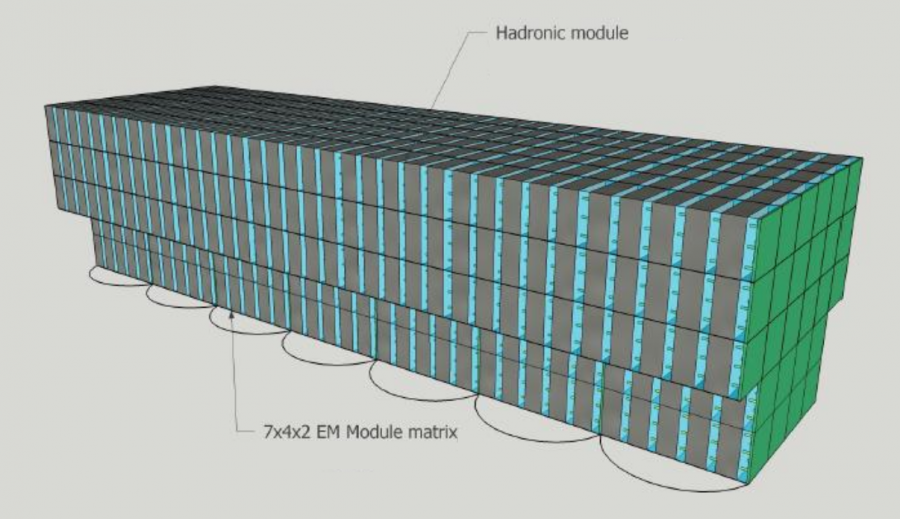}%
} \\
\subfloat[]{%
  \includegraphics[clip,width=0.9\columnwidth]{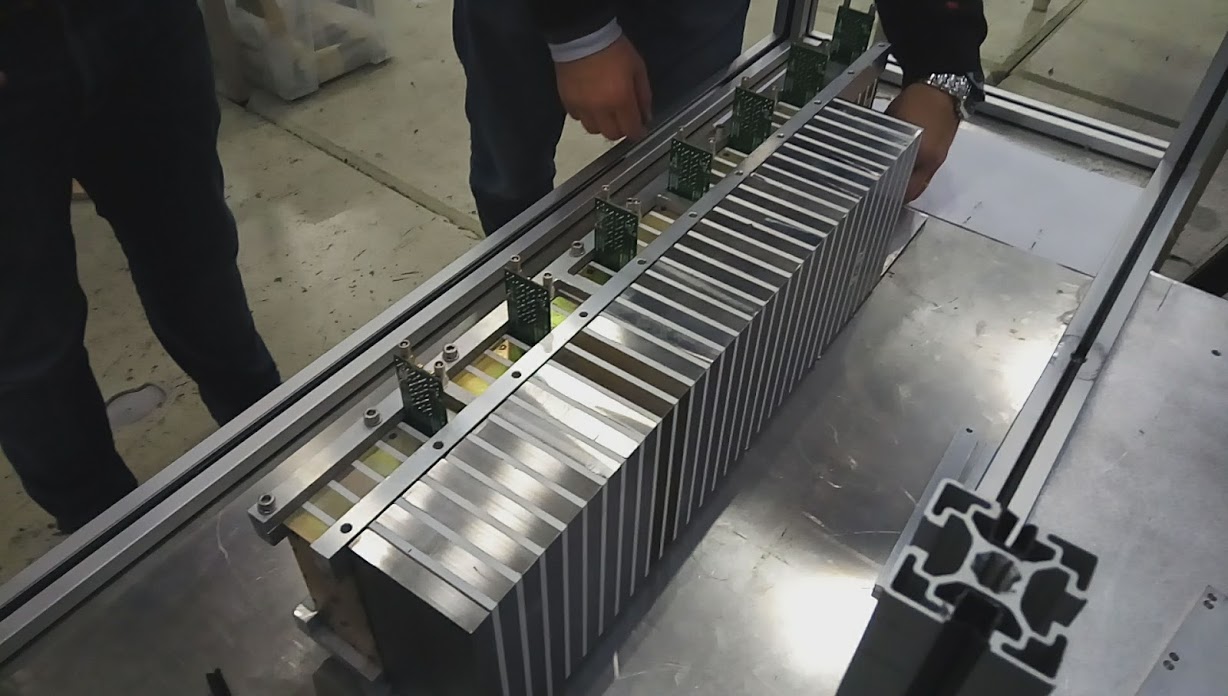}%
}
\end{center}
\caption{\label{fig:layout_calo} (a) Layout of the calorimeter: 7 fine-grained
  modules and, on top, the hadronic module (b) Photo of the
  calorimeter during the installation on the movable platform. The
  calorimeter is rotated by 90$^\circ$ with respect to the layout.  }
\end{figure}

The layout of the calorimeter with the hadronic module on top of the
fine-grained modules is shown in Fig.~\ref{fig:layout_calo}
(a). During the testbeam the calorimeter was rotated by 90$^\circ$ (see
Sec.~\ref{sec:experiment}) and positioned on the movable platform  shown
in Fig.~\ref{fig:layout_calo} (b).

The light transmitted by the fibers is read by 1~mm$^2$ SiPMs with
20~$\mu$m cell size. The sensors are developed by FBK~\cite{FBK}
and are based on the n-on-p RGB-HD technology. Each SiPM hosts 2500
cells in a 1~mm$^2$ square with a fill factor of
66\%~\cite{acerbi}. Each module, corresponding to 8~UCMs, requires the
installation of 9$\times$8=72~SiPMs. The fine grained calorimeter (7
modules) hence hosts 504~SiPMs and the hadron module is read by 162
SiPMs. These SiPMs were produced by FBK from a single wafer and incapsulated
in a chip-scale epoxy package (SMD package) by Advansid
srl~\cite{Advansid}. The V-I response was characterized at the
production site. Since all SiPMs of the calorimeter were produced
starting from the same silicon wafer and in a single lot, the breakdown voltage
is very uniform among the sensors: 28.2$\pm$0.1~V. 

\begin{figure}[!htb]
\centering
\includegraphics[width=0.8\columnwidth]{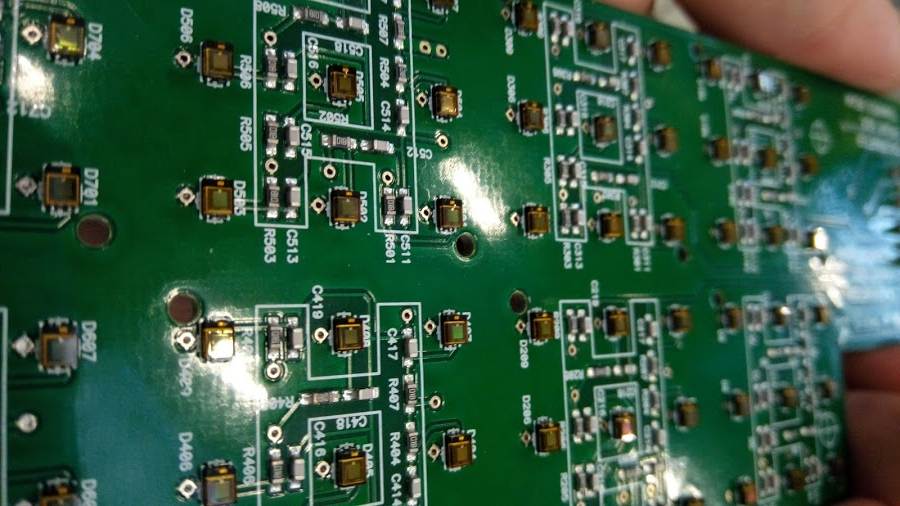}
\caption[]{\label{fig:pcb_zoom} SiPMs embedded in the PCB.  }
\end{figure}

The SiPMs are mounted as standard SMD components on a custom 6-layer
PCB hosting all the sensors belonging to the same module
(Fig.~\ref{fig:pcb_zoom}). The SiPMs belonging to the same UCM are connected
in parallel and read out without amplification through a 47~pF decoupling capacitor.

\begin{figure}[!htb]
\centering
\includegraphics[width=0.9\columnwidth]{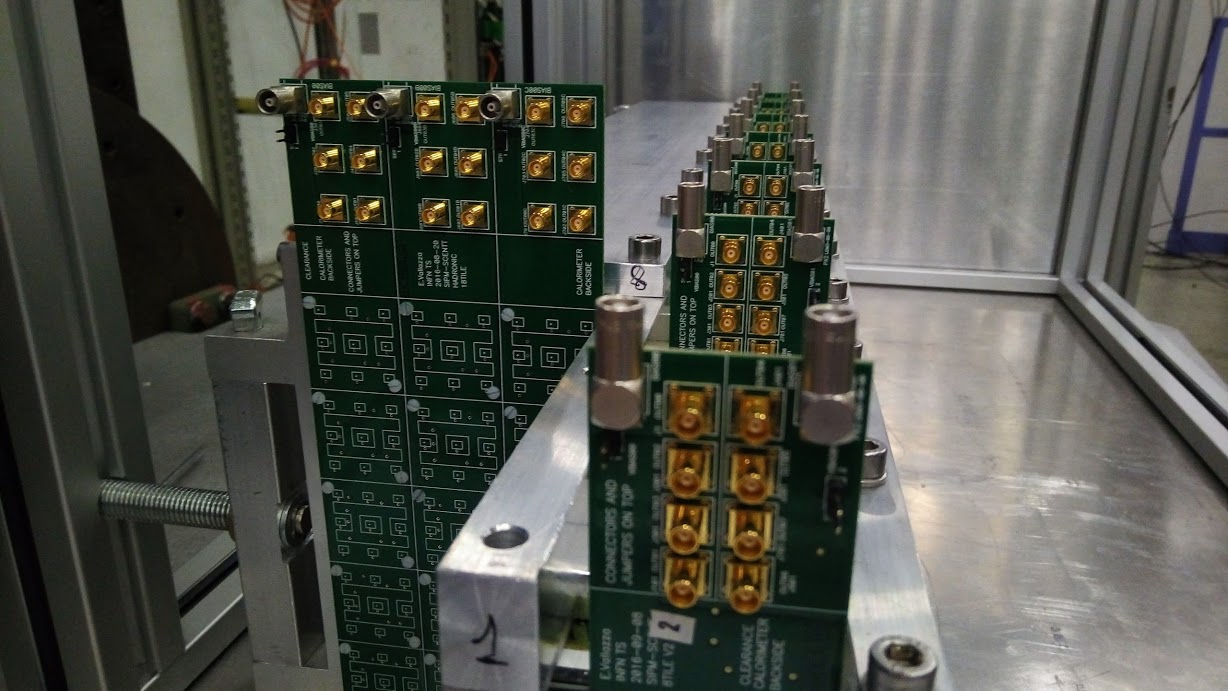}
\caption[]{\label{fig:pcb_mounted} The PCBs mounted in the hadron module
  (left) and in the fine-grained modules (right). }
\end{figure}

The PCB is equipped with a flap hosting 8 MCX connectors to read the
signal of the UCMs (Fig.~\ref{fig:pcb_mounted}). In the PCBs used for
the calorimeter the bias is the same for all SiPMs and it is
distributed by a coaxial cable. During the testbeam all the SiPMs
were biased at 36~V and the equalization among UCMs performed
using the response to minimum ionizing particles (see Sec.~\ref{sec:simulation}).

The hadronic module is read out using a 18$\times$9~cm$^2$ PCB,
corresponding to 18 UCMs.  Unlike the fine-grained modules, however,
the tiles of the UCMs are not optically isolated.  All PCBs are
connected to the back of the modules using PTFE screws positioned
between the PCB and the plastic mask. The SiPMs are aligned to the
fibers in the transverse plane with a precision of 0.1~mm via the
mechanical coupling of the PCB with the plastic mask.

The signals from the UCM are recorded by a set of
8 channel v1720 CAEN~\cite{caen} digitizers (12~bit, 250 MS).
All waveforms are recorded by the DAQ. A reduced dataset is produced for analysis
employing a peak finding algorithm to the waveform data~\cite{Berra:2016thx}.

\section{Test setup in the T9 beamline}
\label{sec:experiment}

\begin{figure}[!htb]
\centering
\includegraphics[width=0.9\columnwidth]{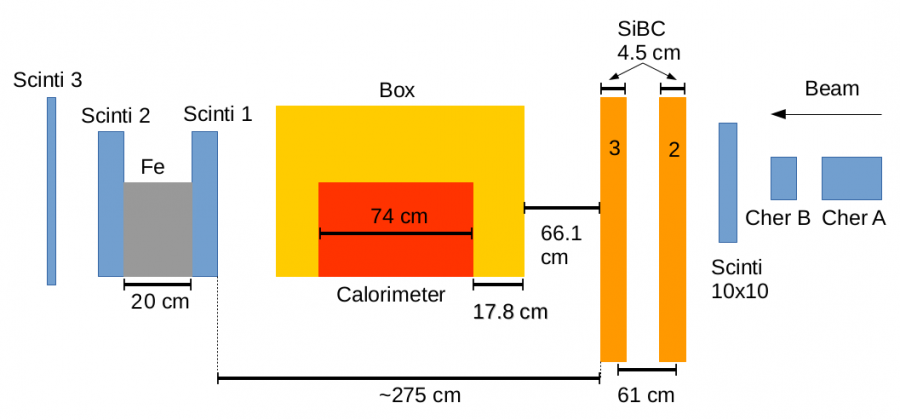}
\caption[]{\label{fig:layout_area} Layout of the instrumentation in
  the experimental area. Detectors include the Calorimeter, two
  Cherenkov counters (Cher A and B), two silicon chambers (SiBC), the
  muon catcher (scint 1 and 2) and the trigger scintillator plane
  (Scinti).}
\end{figure}

\begin{figure}[!htb]
\centering
\includegraphics[width=0.9\columnwidth]{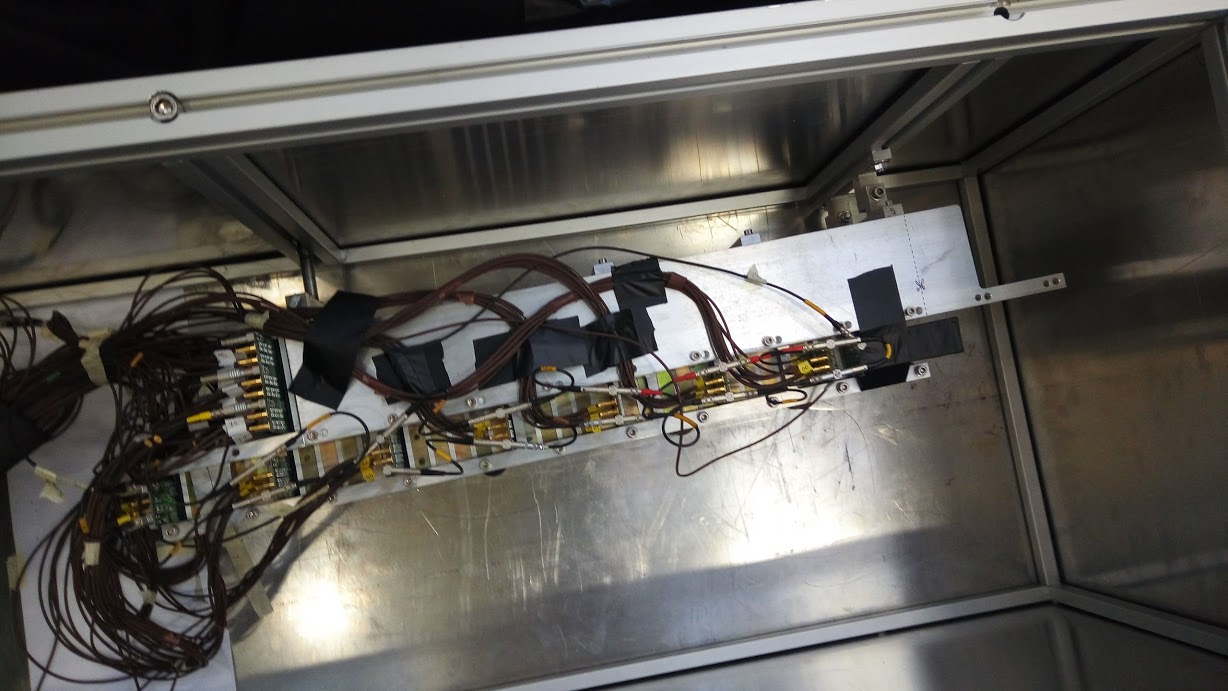}
\caption[]{\label{fig:tilt} The calorimeter positioned inside the
  metallic box and tilted by 100~mrad with respect to the beam axis
  using a threaded bolt (top left of the image).}
\end{figure}

The calorimeter was exposed to electrons, muons and pions at the CERN
PS East Area facility for two weeks in November 2016.  The momentum of
the particles was varied between 1 and 5 GeV, i.e. in the range of
interest for ENUBET. The detector was positioned inside a metallic box
to ensure light tightness and mounted on a platform in the T9
experimental area in front of two silicon strip detectors (see
Fig.~\ref{fig:layout_area}).  During the data taking the calorimeter
was tilted at different angles (0, 50, 100, 200~mrad) with respect to
the beam direction (Fig.~\ref{fig:tilt}).

The silicon detectors~\cite{Prest:2003sy,test_12module} provide track
reconstruction with a spatial resolution of 30~$\mu$m. A pair of
threshold Cherenkov counters filled with CO$_2$ are located upstream
of the silicon detectors. The maximum operation pressure of the
counters is 2.5~bar: they were thus used to separate electrons from
heavier particles ($\mu$ or $\pi$) below 3~GeV.  Between 3~GeV and
5~GeV the two counters were operated at different pressures to
identify electrons, muons and pions.

A $10 \times 10$~cm$^2$ plastic scintillator located between the
silicon and Cherenkov detectors is employed as trigger for the DAQ. Two
pads of plastic scintillator (``muon catcher'') are positioned after
the calorimeter between a 20~cm thick iron shield in order to identify
muons or non-interacting pions.  

Particles in the beamline are obtained from the interaction of the
primary 24 GeV/c protons of the CERN-PS accelerator with a fixed
target. During the test, we employed the T9 ``electron enriched''
target. It consists of an Aluminum tungsten target
(3$\times$5$\times$100~cm$^2$) followed by a tungsten cylinder
(diameter: 10~cm, length: 3~cm). We set the collimators in order to
provide a momentum bite of 1\%.
At 3~GeV the beam composition as measured by the Cherenkov counters is
9\% electrons, 14\% muons and 77\% hadrons.  
We only selected negative
particles in the beamline and the contamination of protons and
undecayed kaons is thus negligible.

The DAQ system is based on a standard VME system controlled by a SBS
Bit3 model 620 bridge, optically linked to a Linux PC-system. The DAQ
is located in the proximity of the calorimeter, inside the
experimental area. Slow control parameters (HV settings for the
Cherenkov counters and the scintillators) and configuration setting
for DAQ (start-stop of the run, quality control) are performed by a
dedicated PC in the Control Room connected to the DAQ PC using a
Gigabit Ethernet link. The acquisition is triggered by the coincidence
of the beam spill (400~ms) and the signal in the plastic
scintillator. The signals from the Cherenkov counters and muon catcher are recorded for each trigger and used off-line for particle identification. Zero suppression in the silicon chambers is performed in
the front-end electronics~\cite{Prest:2003sy}.  During the test,
$\sim$500 particles per spill and 200-500 spills per run were recorded.
The acquisition rate was limited by the waveform digitizers.

\section{Simulation and signal equalization}
\label{sec:simulation}

\begin{figure}[!htb]
\centering
\includegraphics[width=0.9\columnwidth]{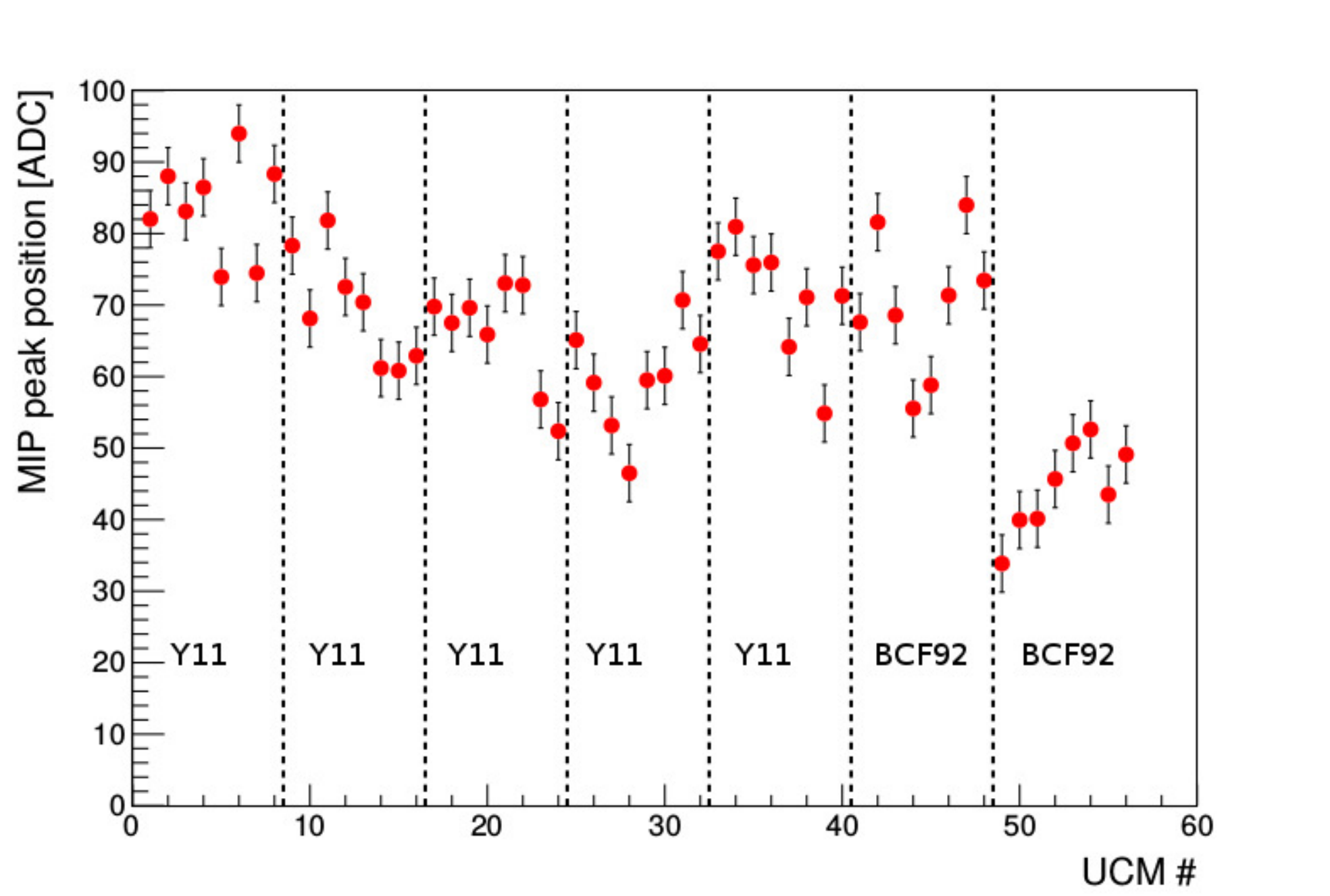}
\caption[]{\label{fig:UCM_response} Signal response to minimum ionizing particles (mip) of each UCM.
The UCMs belonging to the same module are indicated between vertical lines. The WLS fibers used 
for the module are also shown.}
\end{figure}

\begin{figure}[!htb]
\centering
\includegraphics[width=0.8\columnwidth]{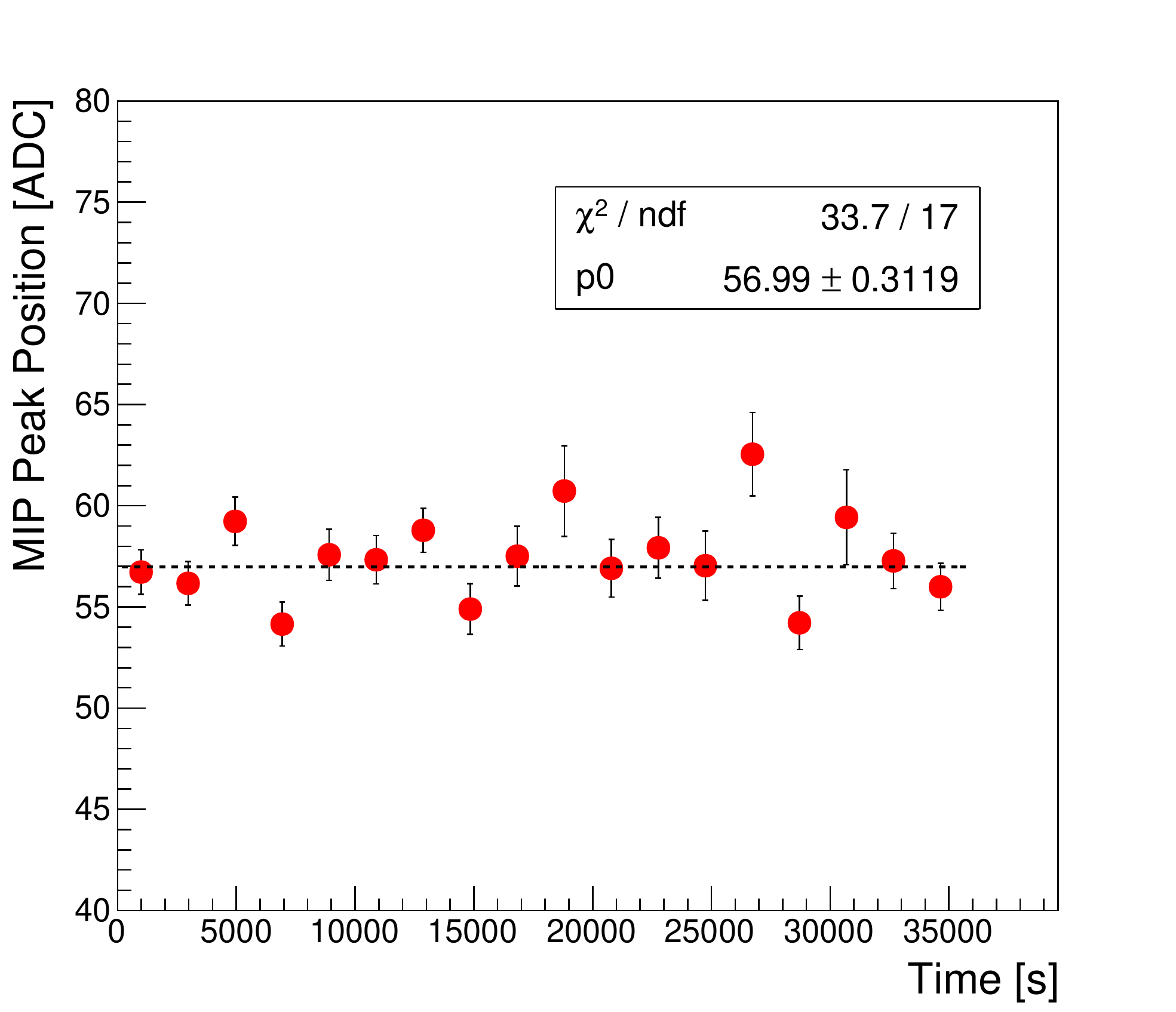}
\caption[]{\label{fig:time_response} Response of one of the UCM to mips as a function of time (in seconds).}
\end{figure}

\begin{figure}[!htb]
\centering
\includegraphics[width=0.8\columnwidth]{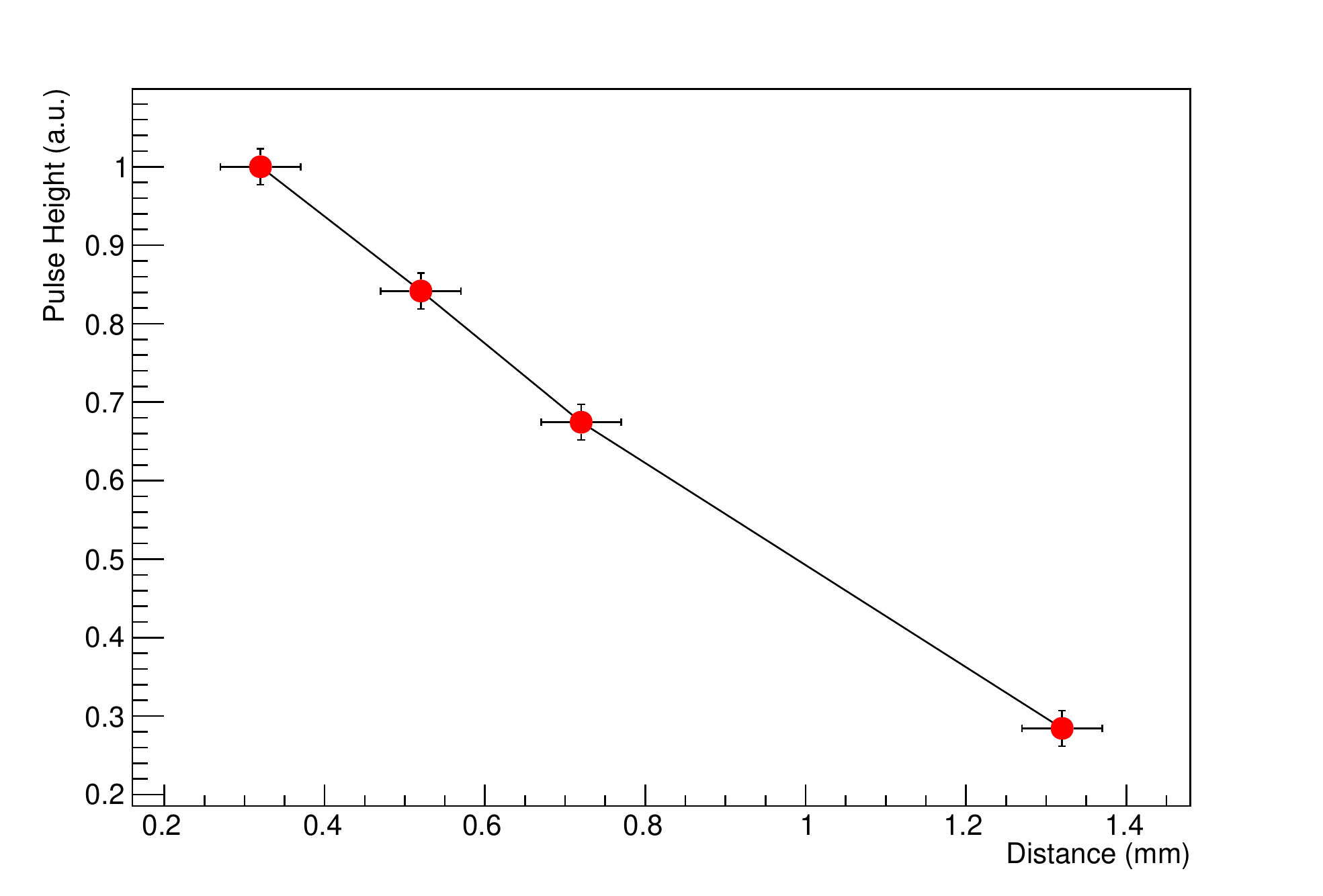}
\caption[]{\label{fig:signal_loss} Pulse height reduction as a function of the distance between the SiPM and the WLS fibers. The closest
point (0.3~mm) corresponds to the thickness of the epoxy package of the SiPM. }
\end{figure}

The signal response to minimum ionizing particles (mip) of each UCM
was measured using a high statistics sample of non-interacting pions
(mip peak) at 9 GeV and checked with a low statistics muon sample at
4~GeV. The distribution of the signal response for each UCM is show in
Fig.~\ref{fig:UCM_response}.  Variation
of the signal response among UCM hosted in the same plastic mask and
SiPM board amounts to 20-30\%. Changes of the response among boards are also
visible both for the UCM equipped with Y11 and
with BCF-92 fibers.  The response is stable in time within the
uncertainty of the position of the mip-like peak
(Fig.~\ref{fig:time_response}) and it is
not correlated with changes in temperature. The dominant contribution
to response non-uniformity is due to misalignment of the fibers
with respect to the SiPM and the use of SiPMs of the same size as the diameter of
the fibers~\cite{Balbi:2014saa}.  Ancillary measurements performed
with single UCMs (Fig.~\ref{fig:signal_loss}) show that
the current mechanical system based on the plastic mask provides a
tolerance of 0.3~mm, which results in a light collection efficiency
reduction of about 25\% and an average number of photoelectrons per
mip of $\sim$40.

The detector response was simulated with GEANT4~\cite{Agostinelli:2002hh,Allison:2006ve,Allison:2016lfl}. The
simulation includes the iron-scintillator tiles, the WLS fibers and
additional material due to the plastic mask and PCB. It does not
include photon generation, transport and the misalignment between
the fibers and the photosensors.  The expected signal in each UCM is
thus proportional to the energy deposit in the scintillator smeared
with the contribution due to photoelectron statistics. The physics list employed
is FTFP\_BERT\_HP~\cite{Allison:2016lfl,Dotti:2011jka}.

\begin{figure}[!htb]
\centering
\includegraphics[width=0.8\columnwidth]{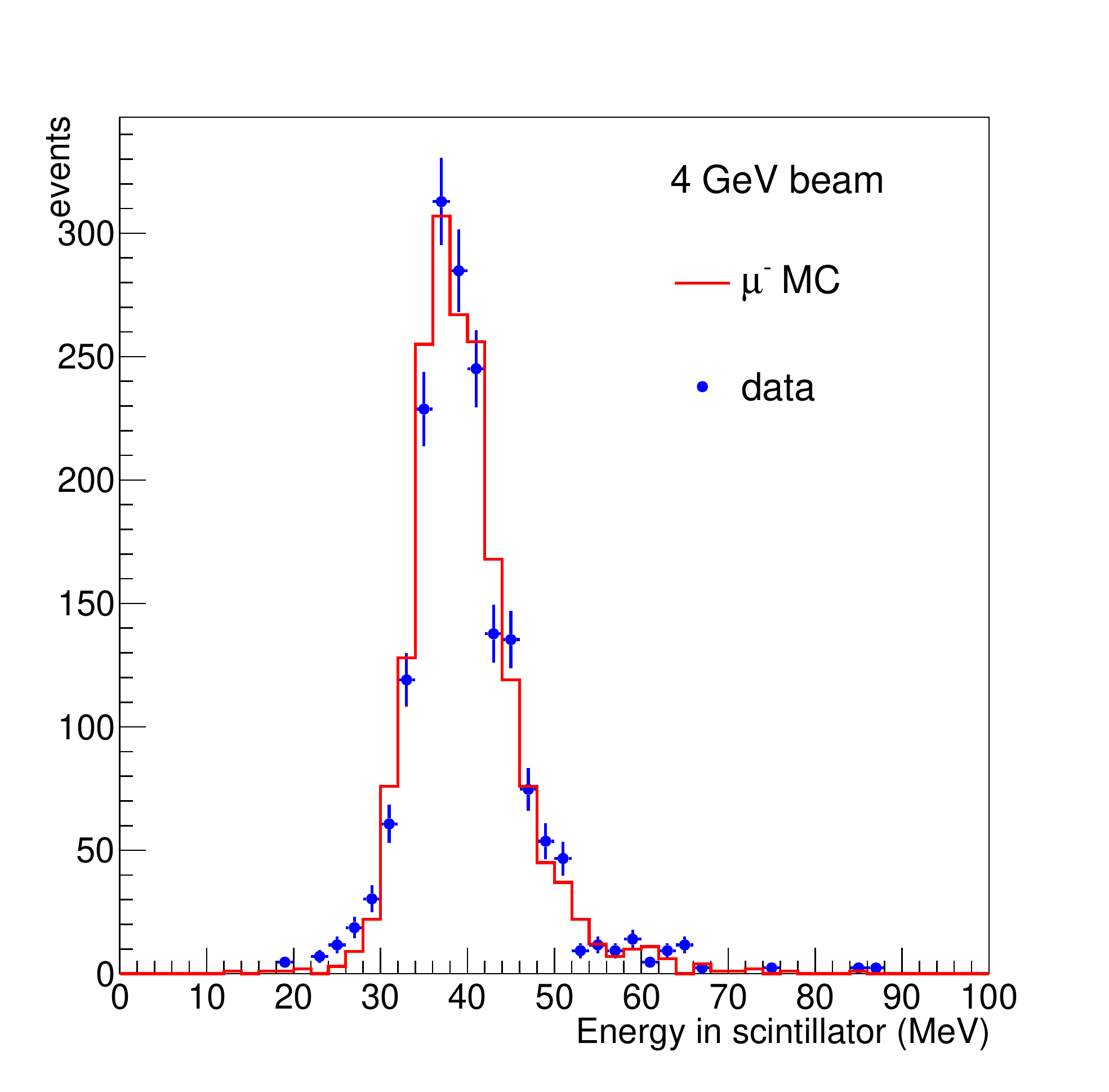}
\caption[]{\label{fig:mip}  Distribution of the energy deposited in the scintillator by 4 GeV
muons impinging on the front face of the calorimeter for data (blue dots) and simulation (red line).}
\end{figure}

The relative response of each UCM is equalized using the mip peak.
Fig.~\ref{fig:mip} shows the response of the detector for 4 GeV
muons impinging on the front face of the calorimeter. 

\section{Response to electrons}
\label{sec:electrons}

The response to electromagnetic showers was evaluated selecting
electrons in the 1-5~GeV energy range. The impact point of the
particles is estimated from the silicon chambers with a precision of
200~$\mu$m.  

The calorimeter under test provides full containment of
electromagnetic showers up to 5~GeV for particles impinging on the
front face and from the lateral side (tilted runs). The
tilted geometry reproduces the operating condition of the calorimeter
in the decay tunnel of neutrino beams, where positrons from $K^+
\rightarrow e^+ \pi^0 \nu_e$ reach the detector with an average angle
of $\sim 100$~mrad~\cite{enubet,Longhin:2014yta}. 

\begin{figure}[!htb]
\centering
\includegraphics[width=0.9\columnwidth]{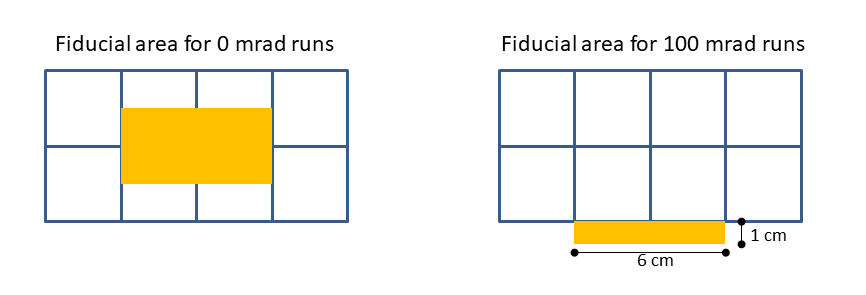}
\caption[]{\label{fig:fiducial} Fiducial areas (gold) selected on the
  front face of the calorimeter projecting the tracks reconstructed
  by the silicon chambers to the uppermost UCM. The left red area
  corresponds to particles impinging in the uppermost UCM from below
  with an angle of 100~mrad}
\end{figure}

The energy and tilt angles of relevance for neutrino physics
applications were tested with dedicated runs at 1,2,3,4 and 5~GeV and
tilts of 0, 50, 100 and 200~mrad. Electrons are selected by the
Cherenkov counters located upstream the silicon detectors. The silicon
chambers are used to select single particles impinging in a
fiducial area with negligible lateral leakage. For particles
impinging on the from face (0~mrad runs), the fiducial area is $6
\times 4$~cm$^2$ (see Fig.~\ref{fig:fiducial}). The fiducial areas for
tilted runs corresponds to particles impinging in the two innermost
UCM from below (see Fig.~\ref{fig:fiducial}). 

\begin{figure}[!htb]
\centering
\includegraphics[width=0.9\columnwidth]{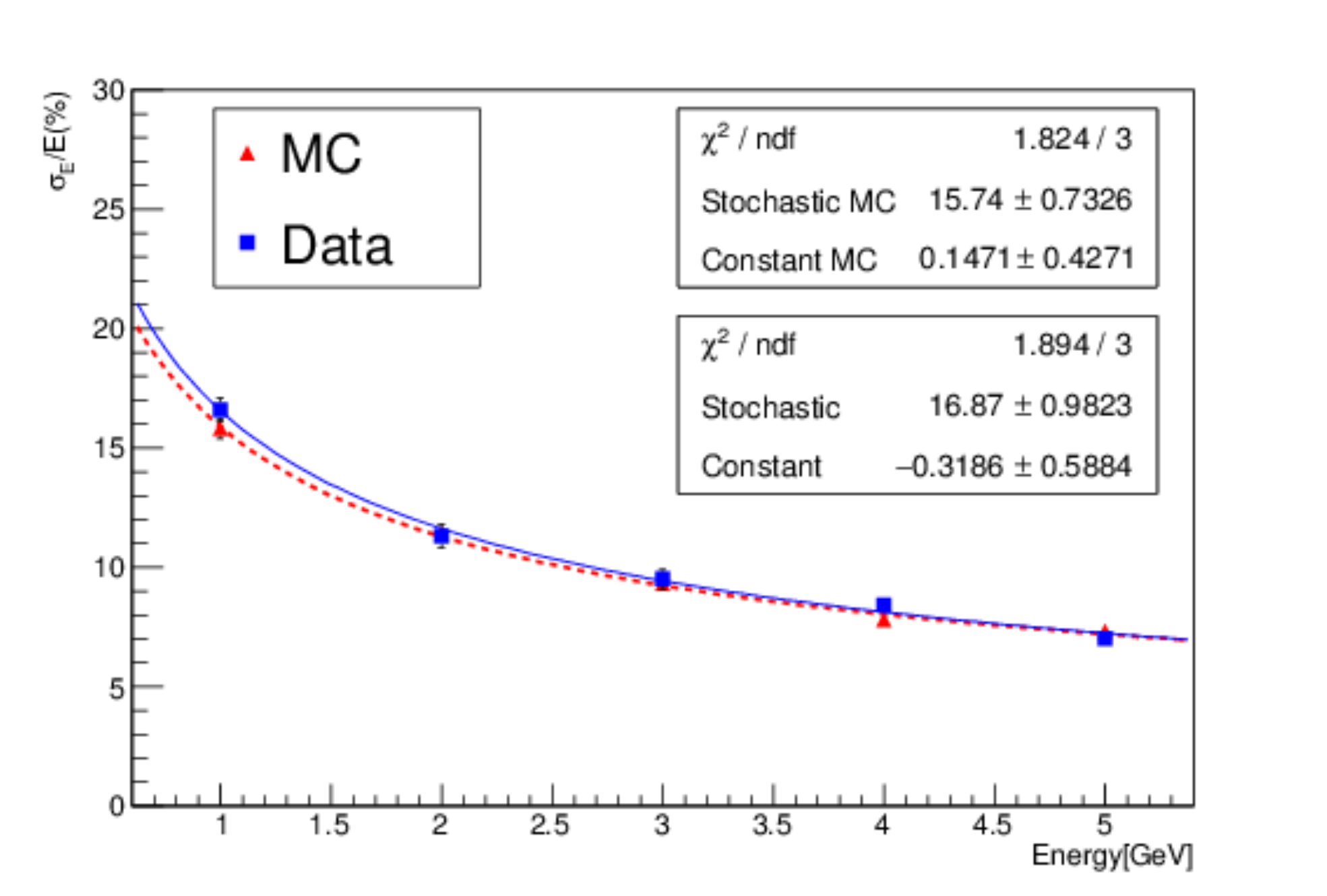}
\caption[]{\label{fig:energy_resolution}  Energy resolution versus beam momentum for particles impinging on the front face (0~mrad run) for data (blue squares)
  and simulation (red triangles). The fit parameters for simulation (MC) and
  data are shown in the top and bottom insets.}
\end{figure}

\begin{figure}[htp]
\begin{center}
\subfloat[]{%
  \includegraphics[clip,width=0.9\columnwidth]{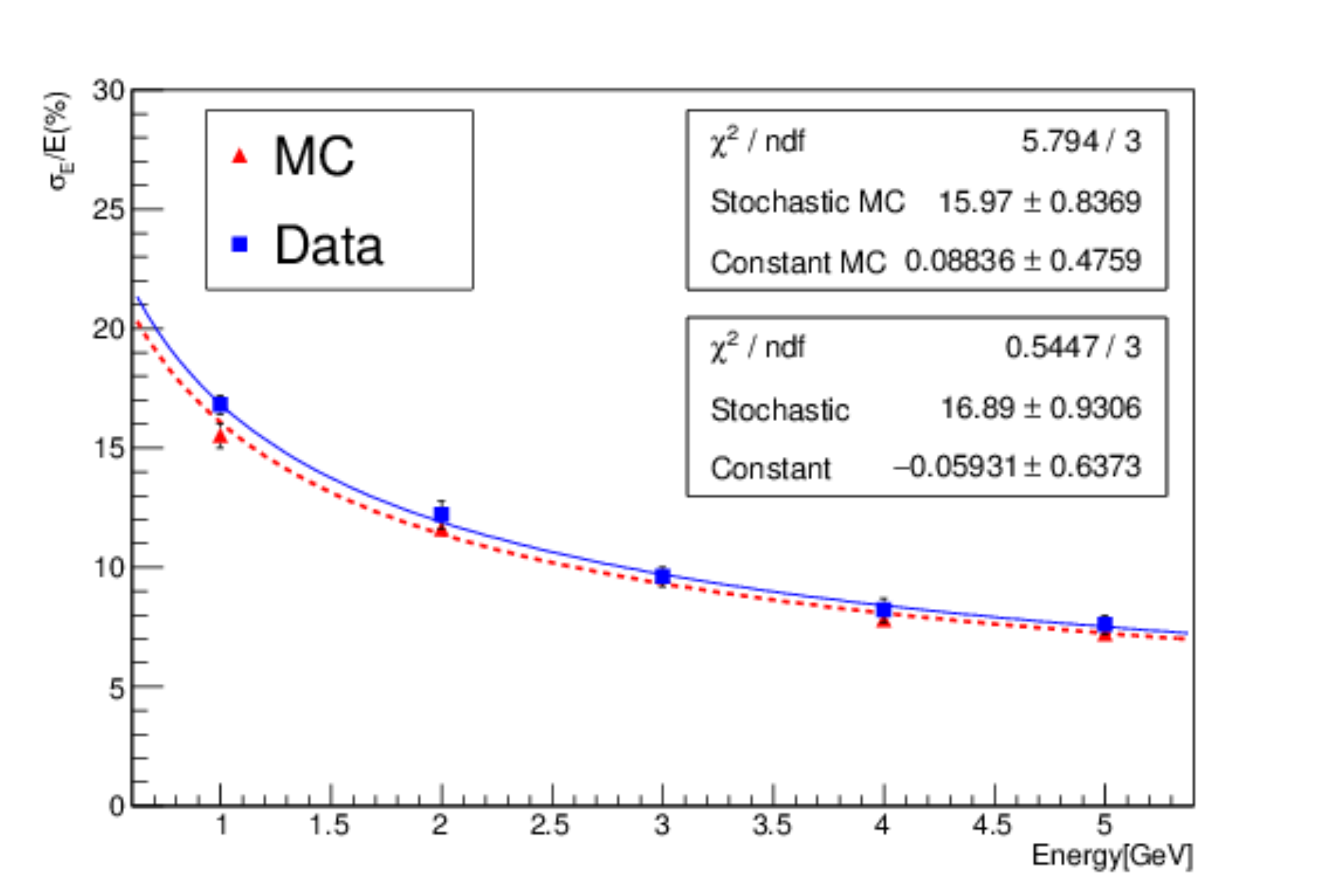}%
} \\
\subfloat[]{%
  \includegraphics[clip,width=0.9\columnwidth]{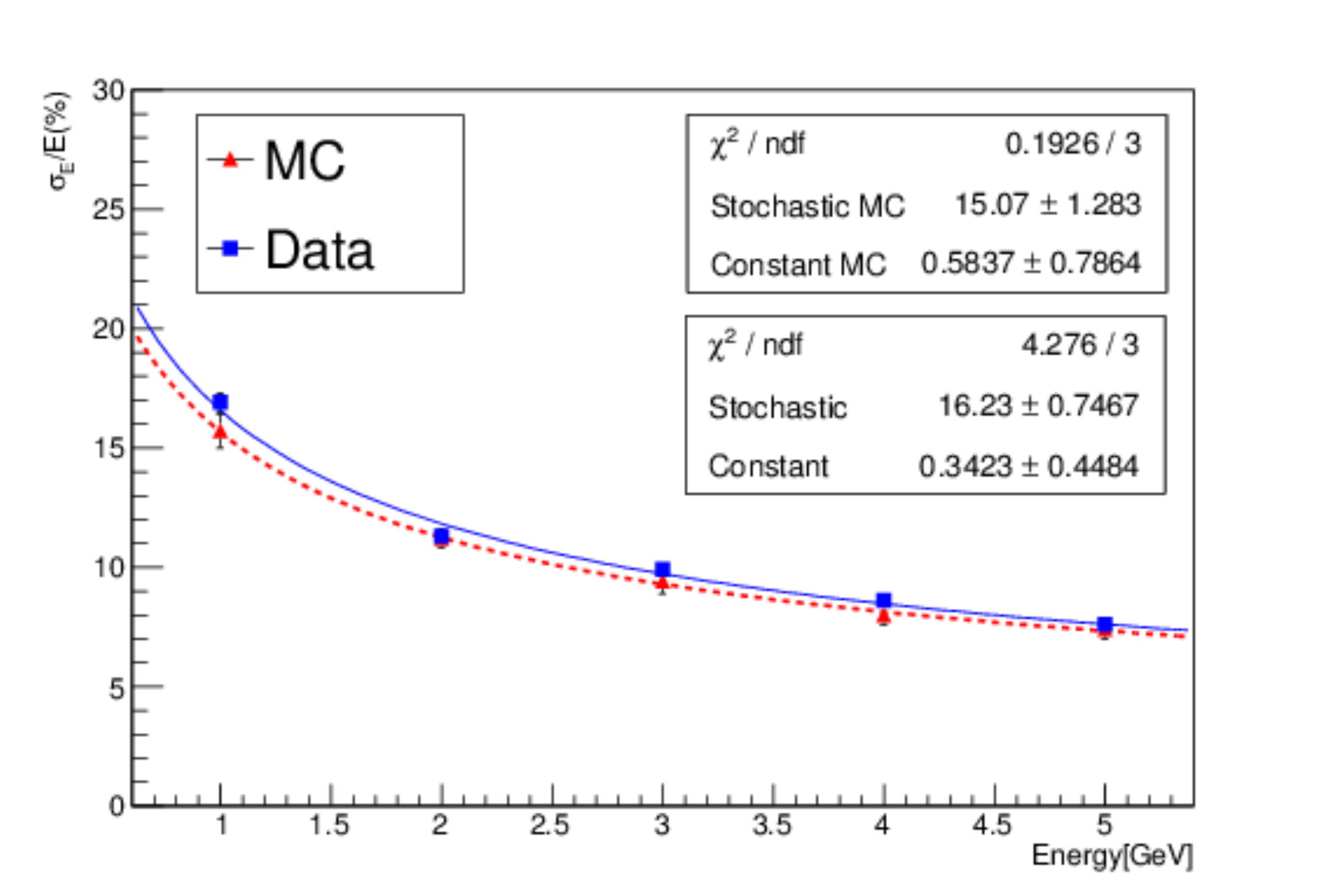}%
}
\end{center}
\caption{\label{fig:energy_resolution_tilt} Energy resolution versus beam momentum for particles impinging at 100~mrad (upper plot) and 200~mrad (lower plot) for data (blue squares)
  and simulation (red triangles). The fit parameters for simulation (MC) and
  data are shown in the top and bottom insets.}
\end{figure}

The energy resolution as a
function of the beam energy for particles impinging on the front face (0~mrad runs) is
shown in Fig.~\ref{fig:energy_resolution} for data (red squares) and
simulation (blue stars). The points are fitted to $\sigma_E/E = S/\sqrt{E
  \mathrm{(GeV)}} \oplus C $, $S$ and $C$ being the sampling
(stochastic) and constant term, respectively. 
Figs~\ref{fig:energy_resolution_tilt}
show the resolution for particles impinging at 100 and 200~mrad.

The 0~mrad runs support the finding of the 12-module prototype,
i.e. that the dominant contribution to the resolution is due to the
sampling term and that the energy response is not affected by the
light readout scheme employed to achieve longitudinal segmentation.
In addition, the analysis of the tilted runs indicates that the
performance of the calorimeter is similar to the 0~mrad run in the
angular range of interest for ENUBET. This behaviour, which is very
relevant for neutrino physics applications, is properly simulated by
GEANT4.

\begin{figure}[htp]
\begin{center}
\subfloat[]{%
  \includegraphics[clip,width=0.8\columnwidth]{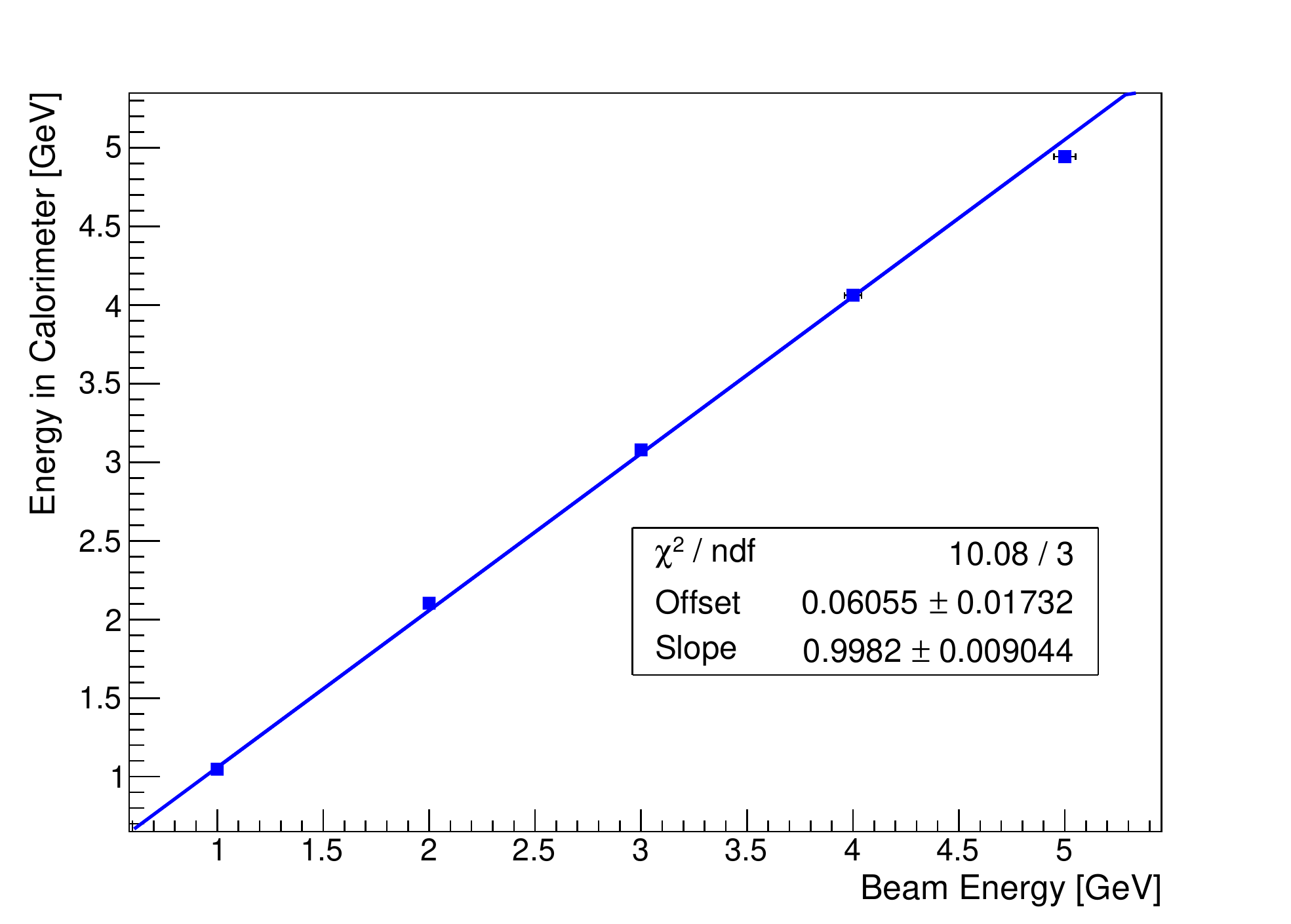}%
} \\
\subfloat[]{%
  \includegraphics[clip,width=0.8\columnwidth]{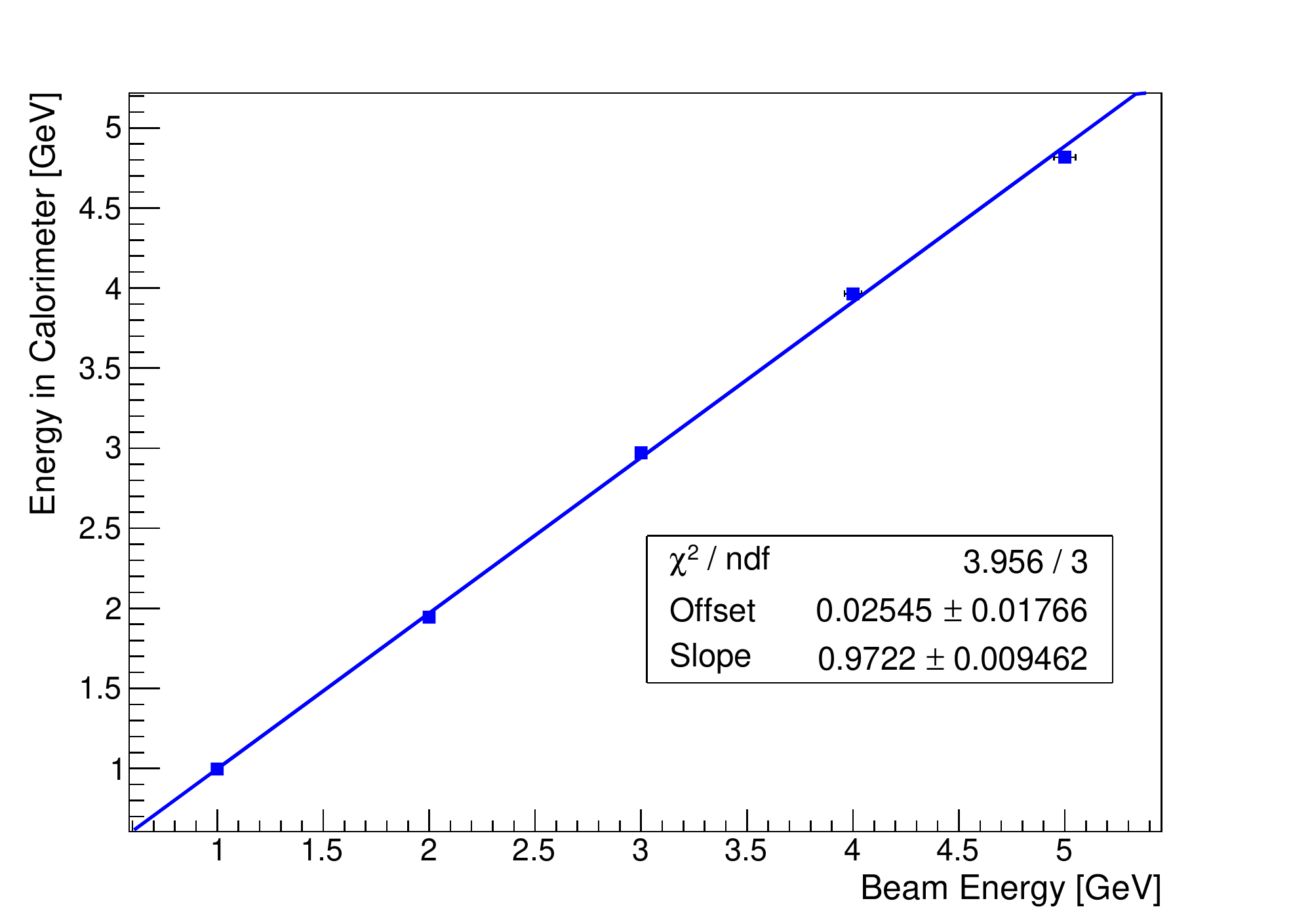}%
}
\end{center}
\caption{\label{fig:linearity} Energy reconstructed in the calorimeter versus beam energy for a 0~mrad (upper plot) and 100~mrad (lower plot) run. The horizontal errors correspond to the momentum bite of the beam.}
\end{figure}

The calorimeter shows linear response in the whole range of interest
within $<$3~\% in both standard (0 mrad) and tilted runs (see
Fig.~\ref{fig:linearity}).
Again, such performance is within the needs for neutrino physics
applications.  Unlike the 12-module prototype, where non-linearities
at high energy could be attributed to longitudinal leakage, the
calorimeter under test provides full containment and the dynamic range
of the SiPM is such that saturation effects are negligible. Residual
non-linearities will be tested with an improved fiber-to-SiPM
connection system and up to a higher energy range.

\section{Response to charged pions}
\label{sec:hadrons}

Longitudinal segmentation is commonly employed in calorimetry for
electron/hadron separation. In fact, a longitudinally segmented
calorimeter is needed in ENUBET for $e^+/\pi^+$ separation in the few
GeV range. In order to validate the ENUBET simulation, a comparison of
the calorimeter response for (negative) pions and electrons has been
carried out. 

\begin{figure}[!htb]
\centering
\includegraphics[width=\columnwidth]{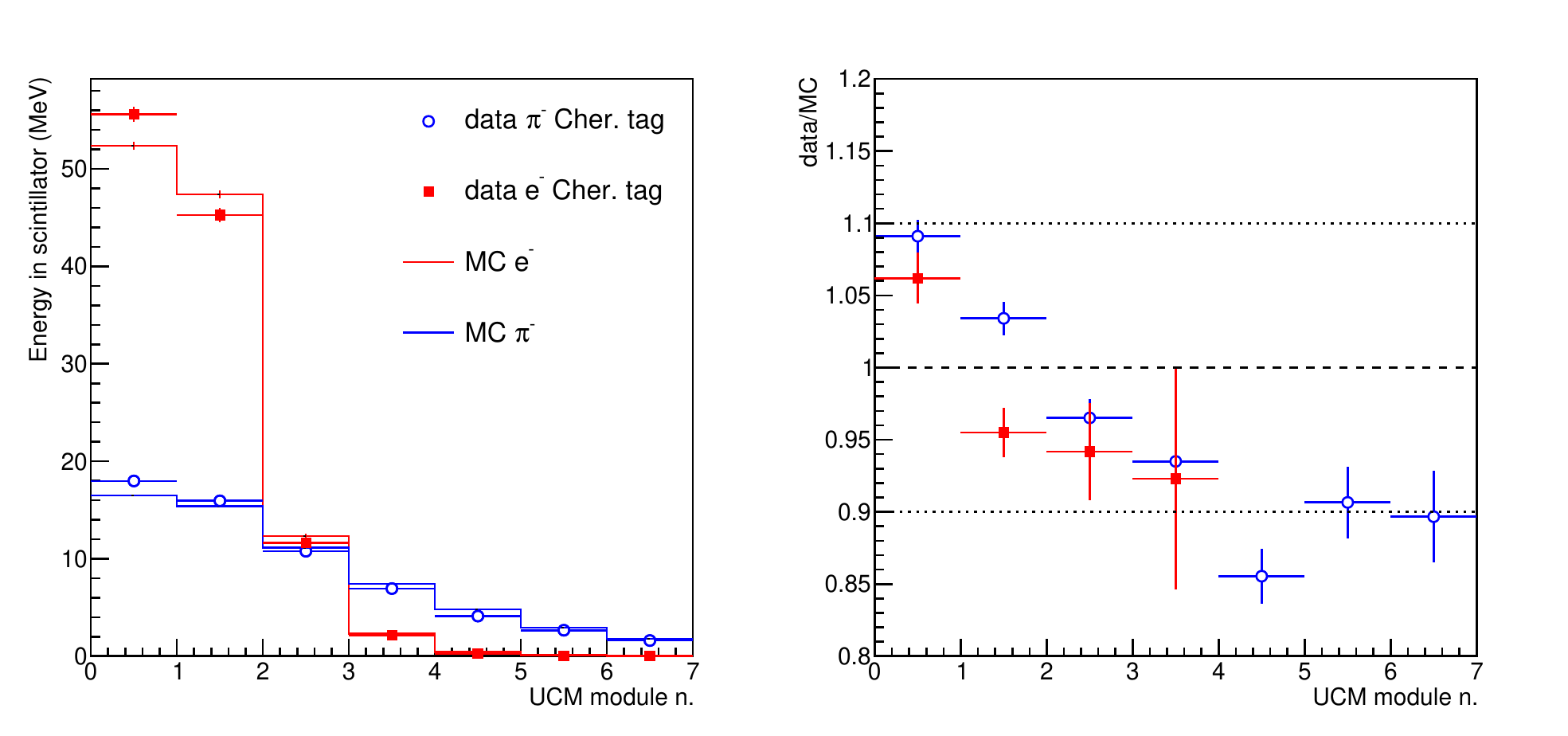}
\caption[]{\label{fig:profile}
(left) Average energy deposited in the scintillator as a function of shower depth for 3~GeV pions and electrons. The depth is expressed in n. of UCM (see text). Each UCM corresponds to 0.44~$\lambda_0$. (right) Energy ratio between data and MC.}
\end{figure}

Fig.~\ref{fig:profile} left (right) shows the average energy deposited
in the scintillator (data/MC ratio) as a function of shower depth for 3~GeV pions and electrons.  
The depth is expressed in number of UCM, i.e. Fig.~\ref{fig:profile} shows  the average energy per event
that is deposited in the UCMs located at a depth $n$ with respect to
the impact point. $n=1$ ($n=7$) corresponds to the most upstream
(downstream) UCM with respect to the beam and each UCM samples 0.44
nuclear interaction lengths ($\lambda_0$).  The whole calorimeter
depth corresponds to $7 \times 4.3 X_0 = 30.1$~$X_0$ and
3.09~$\lambda_0$.

The fiducial area is selected as in
Sec.~\ref{sec:electrons} using the silicon trackers.  In this plot, the energy of the hadronic module is
not employed since this module is not longitudinally segmented.  The
longitudinal profile is reproduced by MC within 10\% for pions and 5\% for electrons. These differences are due
to limitations both in the detector description (photon production and
transport, local variations in the SiPM-to-fiber matching, light
saturation effects~\cite{birks,Akchurin:2014xpa}) and in the
simulation of hadronic showers~\cite{tilecal}.

\begin{figure}[!htb]
\centering
\includegraphics[width=0.8\columnwidth]{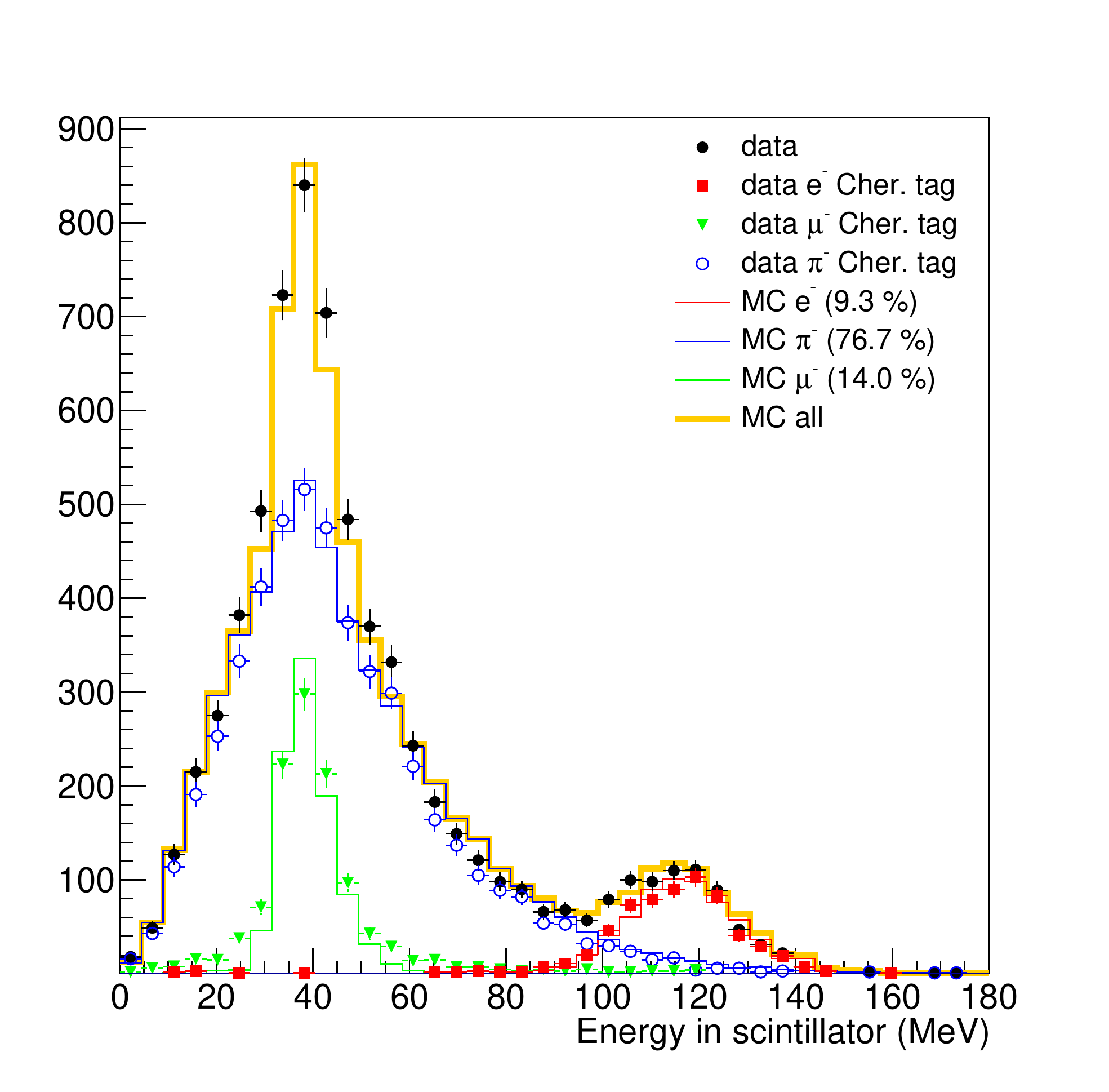}
\caption[]{\label{fig:Etot} Distribution of the energy deposited in
  the scintillator (in MeV) for electrons, muons and pions in a 3~GeV,
  0~mrad run.  The beam composition as measured by the Cherenkov
  detectors is shown in parenthesis. The black dots (thick yellow
  line) is the sum of all particle signals in data (MC).}
\end{figure}

Finally, the energy response of the detector (sum of the signal in all
UCMs) for electrons, muons and pions at 3~GeV is shown in
Fig.~\ref{fig:Etot}. In this plot, the energy scale is determined by
the electron response. The MC response for muons and interacting pions
is re-scaled by 1.10 and 0.91, respectively  to account for the MC
limitations mentioned above. The correction coefficients provide a
good agreement with data in the full energy range (1-5~GeV).

\section{Conclusions}
\label{sec:conclusions}

Fine grained longitudinal segmentation can be achieved in shashlik
calorimeters embedding the SiPMs in the bulk of the calorimeter through
a direct fiber to SiPM connection.  We studied a calorimeter made of
56 modules with a granularity of 4.3~$X_0$ optimized for neutrino
physics applications and we validated its performance at the CERN East
Area facility. The electromagnetic energy resolution - $17\%/\sqrt{E
(\mathrm{GeV})}$ - and linearity - $<$3\% in the 1-5 GeV range - is
appropriate for the needs of ENUBET. No significant changes are
observed in tilted runs up to 200~mrad compared with runs where
particles impinge on the front face.

The fiber-to-SiPM mechanical coupling system needs additional
improvements since it pre\-sen\-tly dominates the non-uniformity of
the response among UCMs. This effect can be corrected equalizing
the UCM response to mip and does not compromize the detector
performance. The MC properly simulates the response to electrons and
muons.  The longitudinal energy profiles of partially contained pions
is reproduced by MC with a precision of 10\%. Further improvements in
the description of the UCM signals call for an amelioration of the
SiPM-to-fiber connection system and a full optical simulation.

\acknowledgments

This project has received funding from the European Union's Horizon
2020 Research and Innovation programme under Grant Agreement
no. 654168 and no. 681647.  The authors gratefully acknowledge CERN
and the PS staff for successfully operating the East Experimental Area
and for continuous supports to the users.  We thank L. Gatignon,
M. Jeckel and H. Wilkens for help and suggestions during the data
taking on the PS-T9 beamline.  We are grateful to the INFN workshops
of Bologna, LNF, Milano Bicocca and Padova for the construction of
detector and, in particular, to D.~Agugliaro, S.~Banfi, F.~Chignoli,
M~Furini, R.~Gaigher, L.~Garizzo, R.~Mazza, L.~Ramina and
F.~Zuffa. Finally, we wish to thank V.~Bonvicini, P.~Branchini and
A.~Lanza for suggestions and support in the design and construction
phase of the experiment.

\end{document}